\documentclass[twocolumn,floatfix,groupedaddress,superscriptaddress,aps,prl,showpacs]{revtex4-1}

\usepackage{graphicx}
\usepackage{amsmath}

\usepackage{amssymb}
\usepackage{amsfonts}

\usepackage{dcolumn}
\usepackage{bbm}

\usepackage{color}

\def\e{\text{e}}

\def\expect#1{\left\langle#1 \right \rangle}

\def\w{\omega}

\begin{document}

\title{Influcence of the nuclear electric quadrupolar interaction on
the  coherence time of hole- and electron-spins confined in semiconductor quantum dots}

\author{J. Hackmann}
\affiliation{Theoretische Physik 2, Technische Universit\"at Dortmund, D-44221 Dortmund, Germany}

\author{Ph.~Glasenapp}
\author{A.~Greilich}
\author{M.~Bayer}
\affiliation{Experimentelle Physik 2, Technische Universit\"at Dortmund, D-44221 Dortmund, Germany}

\author{F.~B.~Anders}
\affiliation{Theoretische Physik 2, Technische Universit\"at Dortmund, D-44221 Dortmund, Germany}

\date{\today}

\begin{abstract}
The real-time spin dynamics  and the spin noise spectra are calculated for p and n-charged quantum dots
within an anisotropic central spin model extended by additional nuclear electric quadrupolar interactions (QC)
and augmented by experimental data studied using identical excitation conditions.
Using realistic estimates for the distribution of coupling constants including an anisotropy parameter,
we show that the characteristic long time scale is of the same order for electron and hole spins strongly determined
by the QC even though the analytical form of the spin decay differs significantly  
consistent with our measurements. 
The low frequency part of the electron spin noise spectrum is approximately $1/3$ smaller than those for hole spins
as a consequence of the spectral sum rule and the different spectral shapes.
This is confirmed by our experimental spectra measured on both types of quantum dot ensembles
in the low power limit of the probe laser.
\end{abstract}





\maketitle

\paragraph{Introduction:}

The promising perspective of combining traditional  electronics
with novel  spintronics  devices lead to
intensive studies of the spin dynamics of a single electron (n) 
or hole (p) confined in a semiconductor quantum dot (QD)
\cite{KhaetskiiLoss2003,Greilich2006,HansonSpinQdotsRMP2007,FischerLoss2008}.
In contrast to  defects in diamonds \cite{JelezkoWachtrup2004,NVreview2006}, such 
QDs may be  integrated into conventional semiconductor
devices. While the strong confinement of the
electronic wave function in QDs reduces the interaction
with the environment and suppresses electronic
decoherence mechanisms, it simultaneously enhances the
hyperfine interaction between the confined electronic spin
and the nuclear spin bath formed by the underlying lattice. 

Generally it is believed 
\cite{Merkulov2002,CoishLoss2004,HansonSpinQdotsRMP2007,FischerLoss2008}
that the hyperfine interaction dominates the spin relaxation
in QDs. The s-wave character of the electron-wave function at the nuclei 
leads to an isotropic  central spin model (CSM) \cite{Gaudin1976}
for describing the electron-nuclear hyperfine coupling, while for
p-charged QDs, the couplings to the nuclear spins can be mapped onto
an anisotropic CSM \cite{FischerLoss2008,Testelin2009}. Since 
the coupling constants for p-charged QDs are reduced
compared to the n-charged QDs  \cite{FischerLoss2008,Testelin2009}, and additionally
a large anisotropy factor 
$\Lambda>1$ suppresses the spin decay of the $S_z$ component
\cite{FischerLoss2008,Testelin2009},
p-charged QDs have been considered as prime candidates
for long lived spin excitations in spintronics applications.

Experimentally, however, there is evidence for comparable spin-decay times  
of the $S_z$  components 
\cite{Crooker2010,PressYamamoto2010,DeGreveYamamoto2011,Varwig2012,Bechtold2014}
in p- and n-charged QDs:
hence the anisotropic CSM provides only an incomplete description of the 
relevant  spin-relaxation processes in such systems.

In this paper, we resolve this puzzle by 
investigating the effect of an 
additional realistic nuclear electric quadrupolar interaction 
term (QC) \cite{Pound1950,*AbragamNMR1961,*Slichter1996}
onto the spin decoherence.
Most of the Ga and As isotopes have a nuclear spin $I=3/2$ which is subject
to a quadrupolar splitting 
in electric field gradients  that occur in 
self-assembled QDs by construction and couple to
the quadrupole moment of the nuclei \cite{Pound1950,*AbragamNMR1961,*Slichter1996}. 
While previously simplified 
assumptions have been made \cite{FlisinskiBayer2010,KuznetsovaBayer2014,Welander2014,Chekhovich2015}, 
or the problem has been mapped on an effective $I=1/2$ model in a random 
magnetic field \cite{Sinitsyn2012} which does not capture the full dynamics,
we have taken into account the proper In dependent anisotropy 
and realistic strain field orientations estimated by a recent microscopic calculation
\cite{Bulutay2012}.
Although the short-time dynamics of p- and n-charged QDs are significantly
different \cite{FischerLoss2008,HackmannAnders2014}, we show that the long
time dynamics is governed by the same time scale set by the quadrupolar 
interactions in agreement with our experimental data presented below.

Over the last decade, an intuitive
picture for the central spin dynamics interacting isotropically with a spin bath 
via hyperfine interaction has emerged. The separation of time scales
\cite{Merkulov2002}  -- a fast electronic precession around an
effective  nuclear magnetic field, and  slow nuclear spin precessions
around the fluctuating electronic spin -- has motivated various
semiclassical approximations 
\cite{Merkulov2002,KhaetskiiLoss2003,Al-Hassanieh2006,ChenBalents2007,Sinitsyn2012,Smirnov2014}
which describe the short-time dynamics of the central spin polarization very well.  
As can be shown rigorously \cite{UhrigHackmann2014} the CSM predicts a finite non-decaying
spin polarization \cite{Merkulov2002,Glazov2012,FaribautSchuricht2013a} whose lower bound
depends on the distribution function of the hyperfine couplings and is only linked
to conservation laws. 
In semi-classical theories \cite{Merkulov2002,Glazov2012} it is given by a third of the initial spin polarization leading
to a large spectral weight at zero-frequency in the spin-noise spectrum.
The absence of such a zero-frequency contribution in experiments
\cite{Crooker2010,Dahbashi2012,LiBayer2012,ZapasskiiGreilichBayer2013}
provides strong evidence that the CSM is incomplete and additional interactions such as QC
play an important role in the decoherence mechanism.

In this work, we have employed a fullly  
quantum mechanical approach, based on a Chebyshev polynomial technique (CET) 
\cite{TalEzer-Kosloff-84,Dobrovitski2003,Fehske-RMP2006}, to
an extended anisotropic spin model.  In order to include QC, we simulate $I=3/2$
nuclear spins.  Within the CET method  the largest accessible time scale or lowest frequency is linearly
connected to the Chebyshev polynomial order.
All technical details can be found in Refs.\  \cite{HackmannAnders2014,Fehske-RMP2006}.

\paragraph{Modelling a quantum dot:}
\label{sec:model}
The dynamics of a single p- and n-charged QD 
is described by the 
Hamiltonian $H$ consisting of three contributions:
\begin{align}
  H &= \frac{g\mu_B B}{\hbar} S^z + H_{\text{CSM}} + H_{\rm QC}. 
  \label{eq hamiltonian_full}
\end{align}
The first term represents an external magnetic field of strength $B$
applied along the growth direction of the QD, which is defined along
the $z$-direction. Furthermore, $\mu_B$ denotes Bohr's magneton, and the occurring
$g$-factor depend on the geometry of the dots and is
different for electrons and holes \cite{Crooker2010}.

The coupling of the central electron or hole spin $\vec{S}$ 
to the nuclear spin bath  can be casted \cite{Testelin2009}
into the anisotropic CSM Hamiltonian $H_{\text{CSM}}$
\begin{align}
  H_{\text{CSM}} =& \sum_{k=1}^{N} A_k \left( S^z I_k^z +
\frac{1}{\lambda} \left( S^x I_k^x + S^y I_k^y \right) \right).
\label{eq hamiltonian}
\end{align} 
$\vec{I}_k$ denotes the
nuclear spin of the $k$-th nucleus, and $N$ is the number of nuclear spins.
The anisotropy parameter $\lambda$ of the spin-flip term  \cite{Testelin2009}
distinguishes between electron ($\lambda=1$) and hole spins,
where $1<\lambda <\infty$ applies depending on the mixture between light and heavy holes.
Due to the enlarged Hilbert space of $2^{2N+1}$ for $I=3/2$, we have restricted
ourselves to $N=10$ in the numerics. This, however, 
reproduces the previous results \cite{HackmannAnders2014}
for $N=20$ nuclear spins with $I=1/2$ in the absence of the QC term.

The energy scale  $A_s = \sum_k A_k$ is expected to be of 
$O(10)\,\mu\text{eV}$ for electrons and approximately one order
of magnitude smaller for holes \cite{Testelin2009}.
The coupling constants $A_k$ are proportional to the squared absolute 
value of the electron or hole
envelope-wave function at the $k$-th nucleus -- 
for details concerning a realistic modelling of the considered
set of $A_k$ entering our numerics see Ref.\ \cite{HackmannAnders2014}.

The  additional quadrupolar term \cite{Pound1950,*AbragamNMR1961,*Slichter1996}
in Eq.\ \eqref{eq hamiltonian_full}
\begin{align}
H_{\rm QC} =& \sum_{k = 1}^{N} q_k \left[ \left( \vec{I}_k \cdot \vec{n}^z_k \right)^2
- \frac{I(I+1)}{3}\right]
 \nonumber\\
 &
  + \frac{q_k \eta}3 \left[ \left( \vec{I}_k \cdot \vec{n}^x_k \right)^2 
-
 \left( \vec{I}_k \cdot \vec{n}^y_k \right)^2 \right]. \label{eq ham_qc}
\end{align}
originates from electric field gradients in self-assembled QDs that  
couple to the nuclear electric quadrupole moment 
and are of crucial importance
for the long-time dynamics of the central spin. 
The coupling constant  $q_k$
is mainly governed by the second order derivative
of the strain induced electric potential $V$  \cite{Pound1950,*AbragamNMR1961,*Slichter1996}.  
The local $z$-direction at the $k$-th nucleus 
is denoted by the normalized orientation vector $\vec{n}^z_k$ which refers to the 
eigenvector corresponding to the largest eigenvalue of  the quadrupolar electric
interaction tensor. The
unit vectors $\vec{n}^{x/y}_k$ complete the local orthonormal basis. 

The asymmetry parameter $\eta = (V_{xx} - V_{yy})/V_{zz}$ 
is commonly neglected in the literature \cite{Dzhioev2007,FlisinskiBayer2010,*KuznetsovaBayer2014,Sinitsyn2012, Welander2014}. 
A recent microscopic calculation of the nuclear electric quadrupolar
couplings \cite{Bulutay2012} in self-assembled  InGaAs QDs, however,
has found values up to $\eta \approx 0.5$ depending on the In 
concentration in the QD. Therefore, we have included 
a finite $\eta=0.5$  in our calculations.

The individual coupling constants $q_k$ are expected
to be up to  $O(1)$neV \cite{Bulutay2012},
but only those  $q_k$ are relevant for the central spin
dynamics where simultaneously $A_k$ is of the same order of magnitude or
larger. We define $A_q = \sum_k q_k$ as a measure of relevant 
total quadrupolar coupling strength which is 
expected to be in the range of $1-100\,\mu$eV
restricting the largest $q_k$ to $q_{\rm max}$.
The ratio $Q_r = A_q / A_s$ determines the relative strength of the QC.

For our simulations, we generate random orientation vectors 
$\vec{n}_k^z$ for each nucleus in our calculation
whose deviation angles are restricted
to $\theta_z \leq 35^\circ$ in accordance to the average deviation
angle $\overline{\theta}_z \approx 25^\circ$ between the growth
direction of the dot and the orientation vectors $\vec{n}_k^z$ for
$\text{In}_{0.4}\text{Ga}_{0.6}\text{As}$ found by Bulutay
\cite{Bulutay2012}. The coupling constants $q_k$
have been generated randomly from a uniform distribution $q_k / q_\text{max}
\in [0.5:1]$.

For $\eta=0$, $H_{\rm QC}$ partially lifts fourfold degenerate  nuclear spin states.
Pinning $\vec{n}^z_k$  to the growth direction, decoherence of the central spin would
be suppressed with increasing $q_k$. A distribution of $\vec{n}^z_k$ due to the inhomogenious
strain fields \cite{Bulutay2012} favors the decoherence. Including
a finite $\eta$
further enhances the decoherence due to the $(S^+)^2+(S^-)^2$ term.

The fluctuations of the transversal
and longitudinal component of the unpolarized nuclear spin bath, 
referred to as Overhauser field, defines the time scale 
$T^*= \lambda /\sqrt{\frac{4I(I+1)}{3} \sum_{k=1}^{N} A_k^2}$
governing the short-time evolution of the central spin
\cite{Merkulov2002, HackmannAnders2014} in the absence of $H_{\rm QC}$. 
We have used this  natural time scale  to define the dimensionless Hamiltonian 
$\tilde{H} = T^* H$.
Two factors in the definition of $T^*$ suggest a longer lifetime for hole spin coherence 
than for electron spins: (i) the coupling constants $A_k$ for holes are typically one
order of magnitude smaller \cite{Testelin2009}  than for electrons,
and (ii) increasing the
parameter $\lambda\ge 1$ to larger values suppresses flips of the
central spin. Both factors enter the time scale linearly, yielding an
expected lifetime increase of a factor $\sim 10\lambda$ for holes compared
to electrons. 
However, when the spin-flip term in $H_{\rm CSM}$ becomes of the order of $H_{\rm QC}$,
this argument fails and the long time decay rate will be strongly influenced by the
QC  for p-doped QDs as we will demonstrate below.

\paragraph{Definition of the spin-noise function:}
The Fourier transformation $S(\w)$ 
of the fluctuation function \\
$S (t) = \frac{1}{2}\left[ \expect{ S^z(t)S^z} +\expect{ S^z S^z(t)} \right] - \expect{S^z}^2$
corresponds to the experimentally measured \cite{Crooker2010,Dahbashi2012,LiBayer2012,ZapasskiiGreilichBayer2013}
spectral power density (see below for experimental details). 
For very small probe laser intensity, 
all expectation values can be calculated using the equilibrium density
operator.  Hence, $S(t)$ is symmetric in time, and
$S(\w)$ is given by
\begin{eqnarray} 
S (\w)&=& \int_{-\infty}^\infty S (t) e^{-i\w t} dt =
\int_{-\infty}^\infty S (t)\cos(\w t) dt \, . \label{s_alpha}
\end{eqnarray} 
From these definitions, we obtain the sum-rule
\begin{eqnarray} 
\int_{-\infty}^\infty \frac{d\w}{2\pi} \, S (\w) &=&
S(0) = \expect{(S^z)^2} -\expect{S^z}^2 
\label{eq sum-rule}
\end{eqnarray} 
for the spin-noise spectrum. In the absence of an
external magnetic field, its value is fixed to $1/4$ for a QD
filled with a single spin. 

Since all experiments are performed in the high-temperature limit,
the inverse temperature $\beta=0$, and
a constant density operator has been used in all numerical calculations.
Then the spin auto-correlation function
$S(t)$ also describes the spin-decay of an initially fully 
polarized central spin \cite{HackmannAnders2014}
interacting with an unpolarized nuclear spin bath, i.\ e.\ $S(t)  = \expect{S_z(t)}/2$.

\begin{figure} [tb] \centering
\hspace{7pt}\includegraphics[width=70mm]{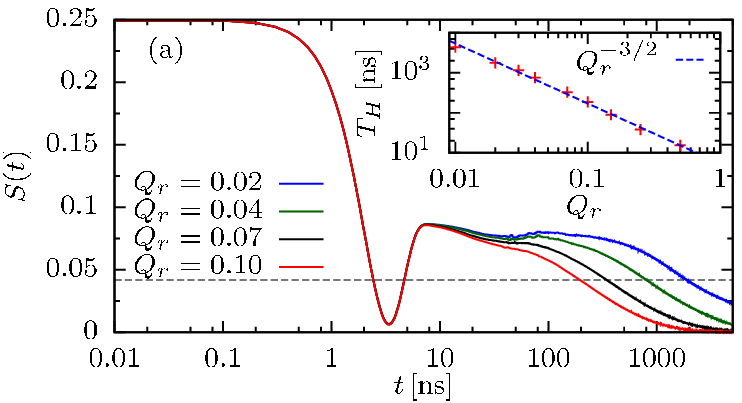}

\includegraphics[width=72mm]{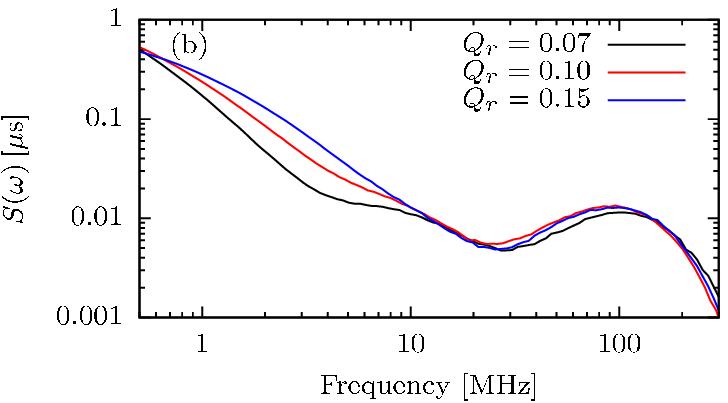}

 \caption{(color online) Panel (a) shows the real time dynamics of a
single electron confined in an InGaAs semiconductor QD for $T^* = 1\,
\text{ns}$ and various values of the parameter $Q_r = A_q / A_s$
without an external field. The inset shows the dependence of the
lifetime $T_H$, which is defined by the crossing of the black dashed
line in panel (a) and $S(t)$ at large times $t \sim O(10^2 - 10^3)
\,\text{ns}$, on the ratio $Q_r$ on a nanosecond scale, which can be
approximated by a power law $\propto Q_r^{-3/2}$. Panel (b) treats
the spectral noise function $S(\omega)$ for various values
of $Q_r$ for electron doped QDs. }
  \label{fig electrons}
\end{figure}

\paragraph{Results:}
For  
various  relative QC strengths  $Q_r$,
Fig.~\ref{fig electrons}~(a) shows
$S(t)$ for electron spins 
($\lambda = 1$.)
The two-stage spin dynamics  is clearly visible:
The initial short-time decay on the scale $T^*$ 
to a plateau of approximately $S(0)/3$ is only  governed by the
Overhauser field \cite{Merkulov2002} and not influenced by QC. Here,
we have used the time scale of $T^*=1$ ns,
see, for example, Ref.\ \cite{Bechtold2014}.
The second stage of the spin-decay is independent of the first 
for small values of $Q_r$, and the  decay is governed by QC. The shape
of our curves agree remarkably with the data of Bechtold et.\ al.~\cite{Bechtold2014}:
$Q_r \approx 0.06-0.1$ seems to be an adequate choice for electrons confined in those InGaAs
QDs.

We have defined a second time scale $T_H$ at which $S(t)$ has dropped to the 
value $S(0)/6$ indicated by the black dashed line 
in Fig.\  \ref{fig electrons} (a) (half the plateau) and have plotted the dependency
of the lifetime $T_H$ on $Q_r$ in the inset. 
$T_H (Q_r)$ approximately obeys a power law $\propto
Q_r^{-3/2}$.

Fig.\ \ref{fig electrons}
(b) shows the spin-noise spectra $S(\w)$
for n-doped QDs for various $Q_r$. The peak 
at around $100\, \text{MHz}$ reflects the short time behavior of $S(t)$ up to
$10\,\text{ns}$ and it is only slightly influenced by the
variation of $Q_r$. Since this peak contains approximately
$2/3$ of the total spectral weight of $S(\omega)$, the signal
of the long time decay for electrons is expected to be a factor
of $3$ smaller than for holes.
As demonstrated in Fig.\ \ref{fig electrons}(a)
the QC mainly impacts the low frequency peak 
corresponding to the long time decay: an increase of $Q_r$ 
broadens the peak width and induces a change of the gradient of
$S(\omega)$ at intermediate frequencies.

\begin{figure} [tb] \centering
\includegraphics[width=72mm]{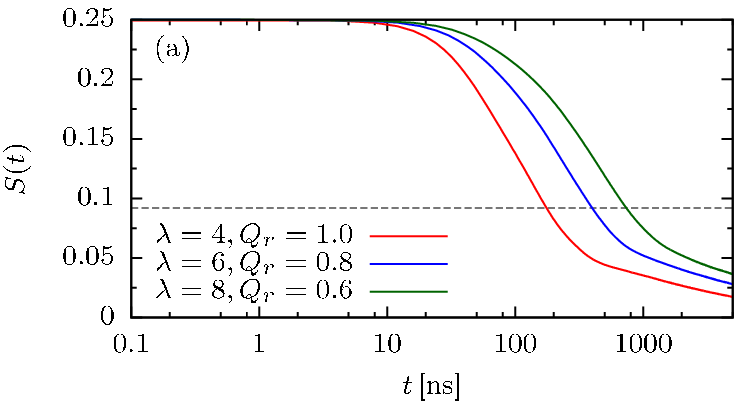}

\includegraphics[width=72mm]{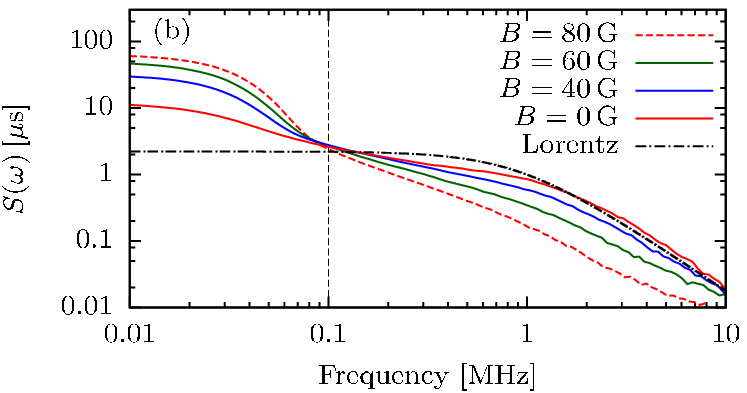}

 \caption{(color online) Spin noise of a hole doped QD in the (a) time
and (b) frequency domain. In panel (a) $S(t)$ is shown for various
combinations of $Q_r$ and the anisotropy parameter $\lambda$
without an external field. Panel (b) shows the spectrum $S(\omega)$
for $\lambda = 4$ and $Q_r = 1$ in a varying external field $\vec{B}$
along the growth direction of the QD. 
For the $B=0$ spectrum we supplemented a Lorentzian (dotted dashed line)
with a width of $g=0.9\, \text{MHz}$.
}
  \label{fig holes}
\end{figure}

Now we focus on p-charged QDs.
Since the overall QC strength $A_q$ does not depend on the doping of the
QD while $A_s$ is decreasing by one order of magnitude when
turning from electrons to holes, the ratio $Q_r$ is increasing
by one order of magnitude at fixed $A_q$. Thus, we expect $T_H$ to decrease by
a factor of $\sim 32$ when turning from electrons to holes. At
the same time $T^*$ is increasing by a factor $10\lambda$, i.~e.\
we expect the lifetime $T_H$ to be of the same order of
magnitude for electrons and holes.

Figure \ref{fig holes}(a) shows  $S(t)$ for
p-charged InGaAs QDs 
for three sets of  parameters $\lambda$
and $Q_r$. For the conversion from the model parameters to the
absolute time scale, we have assumed
a reduction of $A_s$ by a factor of 10 compared to the n-charged case.
For fixed absolute QC parameters $q_k$, $Q_r$ simultaneously increases also by 10,
and, therefore, the absolute values $q_k$  are comparable to those used in
Fig.\ \ref{fig electrons}.
The initial decay due
to the Overhauser field  is suppressed in p-charged QDs by two effects 
that both decrease spin flips of the central
spin on short time scales: (i) the increase of the asymmetry parameter 
$\lambda$ and (ii) the introduced energy splitting to the nuclei due to QC.
Due to the lack of the short-time  spin decay for hole spins, we define
$T_H$  as $S(T_H)=S(0)/\e$, indicated by the black
dashed line in Fig.\ \ref{fig holes}(a). For $\lambda = 4$ and
$Q_r = 1.0$ we have determined the lifetime $T_H = 176\,\text{ns}$
which matches the finding $T_H = 188\,\text{ns}$ for electron spins at
$Q_r = 0.1$ extremely well. For the other parameter sets, 
the lifetimes of $400\,\text{ns}$ ($Q_r=0.8$) and $740\,\text{ns}$ ($Q_r=0.6$) are found,
which are  slightly larger than corresponding electron decay times ($T_H(Q_r) \propto Q_r^{-3/2}$),
but still of the same order of magnitude.

The spin-noise spectrum $S(\omega)$ is shown in Fig.\ \ref{fig holes}(b)
for various external longitudinal magnetic field strengths $B$, $\lambda = 4$ and $Q_r = 1.0$
\footnote{
For the other two sets of parameters depicted in Fig.\ \ref{fig holes}~(a)
qualitatively the same results are found.
}.
The calculated 
$S(\w)$  corresponds to recent measurements \cite{LiBayer2012} 
and a nice agreement between our theory and the
experiments is found: for increasing $B$ the spectral weight,
fulfilling the sum rule (\ref{eq sum-rule}), is shifted from large
to small frequencies. As a consequence the gradient of $S(\omega)$
in the intermediate frequency regime $\omega\sim O(0.1)\, \text{MHz}$
is increasing, which is referred to as a shift from an approximately
Lorentzian lineshape for $B=0$ to a $1/f$ noise with increasing $B$ 
as reported in Ref.\ \cite{LiBayer2012}. Unfortunately, the
resolution of our numerical investigations is limited to $\sim0.1\, \text{MHz}$ 
for this parameter regime requiring already $6000$ Chebychev polynomials.
The linewidth of the added Lorentzian (dotted dashed line)
at half width half maximum 
is $0.9\, \text{MHz}$, corresponding to the observed
lifetime $T_H = 176$ ns. 
Note that for the parameter set
$\lambda=6$ and $Q_r = 0.8$ the corresponding linewidth is
$400\,\text{kHz}$, which matches the experimental findings of
Ref.\ \cite{LiBayer2012}. 

\begin{figure} [tb] \centering

\includegraphics[width=70mm]{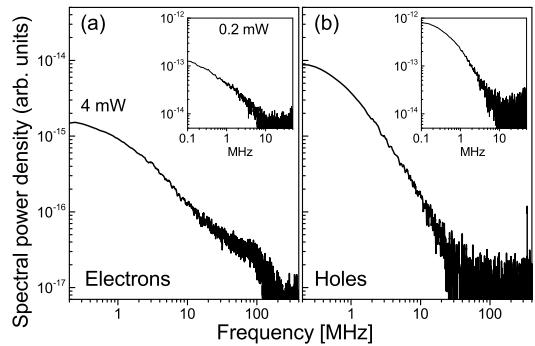}

 \caption{(color online) 
Measured spin noise spectra at zero external magnetic
field for electron (a) and hole spins (b) in ensembles of single charged
QDs, measured around 890 nm laser excitation wavelength at 4 mW power
level. The spectra are measured using a $650 \, \text{MHz}$ bandwidth receiver
(1607-AC, NewFocus). Insets show measurements at 0.2 mW laser power
using a sensitive $100\, \text{MHz}$ detector (HCA-S, Femto), confirming the
identical zero frequency peak width. 
}
  \label{fig experiment}
\end{figure}
 
For further comparison with our calculation, Fig.~\ref{fig experiment}
shows experimentally
measured spin noise spectra at a temperature of 5 K. The experiments were
performed on (In,Ga)As quantum dot ensembles of similar dot density, in
one case on average doped by a single electron per dot, in the other
case by a single hole \cite{Crooker2010,ZapasskiiGreilichBayer2013}.
The samples were studied using identical excitation conditions. The
linearly polarized light beam of a single frequency laser was tuned to
the ground state transition energy maximum \cite{Crooker2010}. The laser
power was reduced to 4 mW focused into a spot of $100\,\mu\text{m}$ diameter, giving
a good signal to noise ratio in 10-20 minutes of accumulation time, while
simultaneously minimizing the laser excitation impact \cite{LiBayer2012}. The
noise spectra are taken by a real time FFT using a FPGA module
\cite{Crooker2010} and the spin-component is retrieved from the noise
background by interlacing  the data at zero and 250 mT
magnetic field applied
in Voigt direction. At 250 mT the peaked contribution to the noise due
to spin precession is shifted out of the measured spectral range.

The comparison of the electron and hole spin noise spectra in
Figs.~\ref{fig experiment}(a) and (b) with the calculations reveals
that the theory qualitatively correctly predicts the shape and
widths of the spin-noise spectra. In particular, the following
features are worth noting: (i) The electron spin noise shows an
additional peak around $100 \, \text{MHz}$ unveiling the electron's precession
in the frozen Overhauser field \cite{Merkulov2002}, as also present
in Fig.\ \ref{fig electrons}(b). (ii) Since  $S(\w)$ must obey the
sum-rule \eqref{eq sum-rule}, the low-frequency spectral weight of
$S(\w)$ for n-charged QD is only about $1/3$ of those for holes. A
Lorentzian fit to the low frequency components ($f <
35 \,\text{MHz}$) of the experimental data confirms this difference
in the amplitudes. (iii) a spin correlation time of the same order
of magnitude in the long-time range for electrons and hole spins, as
predicted by the theory. In the experiment this time is on the 
order of $400\, \text{ns}$, as estimated from the peak width at low
frequencies.

\paragraph{Summary:}
We have compared the impact of the hyperfine interaction on the spin
coherence in n- and p-charged QDs, including the nuclear quadrupolar
electric interaction generated by the strain fields, which provides
an additional decoherence mechanism acting equally for n- and
p-charged QDs. This mechanism is sufficient to explain the very
similar long-time decay time $T_H$ of n- and p-charged QDs. On the
other hand, the different coupling of electron and hole spins in the
central spin part of the Hamiltonian leads to significant deviations
in the short term dynamics, most prominently evidenced by the
electron spin precession about the nuclear magnetic field.

The samples have been provided by D. Reuter und A.D. Wieck,  
Bochum University, Germany.
This work has been supported by the DFG and the RFBR through the TRR 160.



%

\end{document}